\documentclass[twocolumn,aps,prd]{revtex4}
 \usepackage{graphicx,amssymb}
 \usepackage[utf8]{inputenc} 
 
\usepackage[urlcolor=blue,colorlinks,breaklinks=true]{hyperref}

\PassOptionsToPackage{obeyspaces,spaces,hyphens}{url}
 
 \usepackage{amsmath}

\begin{document}
 	\def\half{{1\over2}}
 	\def\shalf{\textstyle{{1\over2}}}
 	
 	\newcommand\lsim{\mathrel{\rlap{\lower4pt\hbox{\hskip1pt$\sim$}}
 			\raise1pt\hbox{$<$}}}
 	\newcommand\gsim{\mathrel{\rlap{\lower4pt\hbox{\hskip1pt$\sim$}}
 			\raise1pt\hbox{$>$}}}

\newcommand{\be}{\begin{equation}}
\newcommand{\ee}{\end{equation}}
\newcommand{\bq}{\begin{eqnarray}}
\newcommand{\eq}{\end{eqnarray}}
 	
\title{Probing gravity at sub-femtometer scales through the pressure distribution inside the proton}
 	 	
\author{P.P. Avelino}
\email[Electronic address: ]{pedro.avelino@astro.up.pt}
\affiliation{Instituto de Astrof\'{\i}sica e Ci\^encias do Espa{\c c}o, Universidade do Porto, CAUP, Rua das Estrelas, PT4150-762 Porto, Portugal}
\affiliation{Centro de Astrof\'{\i}sica da Universidade do Porto, Rua das Estrelas, PT4150-762 Porto, Portugal}
\affiliation{Departamento de F\'{\i}sica e Astronomia, Faculdade de Ci\^encias, Universidade do Porto, Rua do Campo Alegre 687, PT4169-007 Porto, Portugal}
 	
\date{\today}
\begin{abstract}
 
Recently, a measurement of the pressure distribution experienced by the quarks inside the proton has found a strong repulsive (positive) pressure at distances up to 0.6 femtometers from its center and a (negative) confining pressure at larger distances. In this paper we show that this measurement puts significant constraints on modified theories of gravity in which the strength of the gravitational interaction on microscopic scales is enhanced with respect to general relativity. We consider the particular case of Eddington-inspired Born-Infeld gravity, showing that strong limits on $\kappa$, the only additional parameter of the theory with respect to general relativity, may be derived from the quark pressure measurement ($|\kappa| \lsim 10^{-1} \, {\rm m^5 \, kg^{-1} \, s^{-2}}$). Furthermore, we show how these limits may be significantly improved with precise measurements of the first and second moments of the pressure distribution inside the proton. 
		
\end{abstract}

\maketitle
 	
\section{Introduction}
\label{sec:intr}

Achieving a detailed understanding of the precise nature of quark confinement is one of the main goals of modern particle physics (see \cite{Polyakov:2018zvc} for a recent review).  In a recent work \cite{Burkert:2018bqq} the pressure distribution experienced by the quarks inside the proton has been measured using deeply virtual Compton scattering \cite{1997PhRvL..78..610J,1997PhRvD..55.7114J}. The experimental data \cite{Jo:2015ema}, obtained at an electron beam energy of  $6 \, \rm GeV$, provided evidence of a strong repulsive (positive) pressure at distances up to 0.6 femtometers from the center of the proton and a (negative) confining pressure at larger distances. This confining pressure is responsible not only for the stability of the proton, but also of most of the elements of the periodic table (which account for the almost entire baryonic content the Universe). The quark pressure measurement also revealed an average pressure near the center of the proton of about $10^{35} \, \rm Pa$, which is larger than the estimated pressure at the heart of neutron stars \cite{Ozel:2016oaf}. Moreover, a significant improvement in the precision of the results is expected with new experiments at $12 \, \rm GeV$ performed at Jefferson Lab \cite{Burkert:2018nvj}.

In the present paper we address the question of whether these experiments can provide interesting constraints on the properties of gravity on femtometer and sub-femtometer scales.  In order to play a relevant role inside the proton the gravitational field strength on such scales would have to be larger, by many orders of magnitude, than the general relativity prediction. This is not a problem on itself, since viable modified theories of gravity in which the gravitational interaction on microscopic scales can be enhanced with respect to general relativity have been proposed (see \cite{Berti:2015itd,BeltranJimenez:2017doy} for recent reviews).

In this paper we shall focus on Eddington-inspired Born-Infeld (EiBI) gravity \cite{Banados:2010ix} (see also \cite{Deser:1998rj, Vollick:2003qp,Vollick:2005gc,Vollick:2006qd}) a formulation of gravity inspired by Born-Infeld non-linear electrodynamics \cite{Born:1934gh} and its solution to the problem of the divergent self-energy of point charges--- EiBI gravity is a particular example of a wide class of Ricci-based metric-affine theories of gravity \cite{BeltranJimenez:2017doy,Latorre:2017uve}. Although EIBI gravity is equivalent to Einstein’s general relativity in vacuum, inside matter the strength of the gravitational field can be much larger than in general relativity. This is due to an effective gravitational pressure \cite{Avelino:2012ge,Delsate:2012ky}, which could potentially be responsible for the avoidance of some astrophysical and cosmological singularities that would be predicted to exist in the context of general relativity \cite{Banados:2010ix,Avelino:2012ue,Bouhmadi-Lopez:2013lha,Bouhmadi-Lopez:2014jfa}. In EiBI gravity strong deviations from general relativity may arise at the extremely high densities attained in the early universe \cite{Banados:2010ix,Avelino:2012ue,Scargill:2012kg}, in the interior of compact astrophysical objects such as black holes \cite{Olmo:2013gqa,Sotani:2014lua, Wei:2014dka, Jana:2015cha, Avelino:2015fve,Avelino:2016kkj} and neutron stars \cite{Pani:2011mg,Pani:2012qb,Harko:2013wka}, or even inside subatomic particles such as the proton \cite{Avelino:2012qe}.

The outline of this paper is as follows. In Sec. \ref{sec:gr} we start by considering the role of the pressure inside stable compact objects in the Newtonian limit of general relativity. There, we provide an estimate of the average pressure as a function of the gravitational potential at the surface of compact spherical symmetric objects in hydrostatic equilibrium,  discussing the particular case of the proton. In Sec. \ref{sec:eibi} we briefly describe the EiBI theory of gravity, giving particular relevance to the concept of effective gravitational pressure which emerges naturally in this theory. In Sec. \ref{sec:proton} we study the constraints on modified theories of gravity that may be obtained from the recent measurement of the pressure distribution inside the proton, paying special attention to the case of EiBI gravity. We also discuss the improvement of the constraints which will result from precise measurements of the first and second moments of the pressure distribution inside the proton. We then conclude in Sec.  \ref{sec:conc}.

In this paper we adopt a metric signature $(-,+,+,+)$. The Einstein summation convention will be used when a greek or a latin index variable appears twice in a single term, once in an upper (superscript) and once in a lower (subscript) position. Greek indices and the latin indices $i$ and $j$ take the values $0,...,3$ and $1,...,3$, respectively. We will use angle brackets to denote averages. Except when stated otherwise, we shall use units with $c = 8\pi G = 1$.

\section{Average pressure inside compact objects in general relativity}
\label{sec:gr}

The average pressure $\langle p \rangle$ inside stable compact objects with a negligible self-induced gravitational field is equal to zero. This condition, also known as the von Laue condition, is discussed in  depth in \cite{Polyakov:2018zvc}  --- see also \cite{Avelino:2018qgt,Avelino:2018rsb} for a detailed derivation of the von Laue condition for generic matter fluids and of its relation to the form of the Lagrangian of a fluid composed of solitonic particles. In the case of a spherically symmetric object of radius $r_*$ and mass $M_*$, the vanishing average pressure condition may be written as
\be
\langle p \rangle = \frac{\int_0^{r_*} p(r) r^2 dr}{\int_0^{r_*} r^2 dr}=0\,, \label{averagep}
\ee
where $r$ is the radial coordinate and $p \equiv  {{{T}^{i}}_{i}}/{3}$ is the proper isotropic pressure.

Let us now consider a case in which the self-induced gravitational field cannot be neglected. In the Newtonian limit of general relativity, a non-relativistic spherically symmetric compact object made of a perfect fluid in hydrostatic equilibrium (e.g. a star) satisfies the equation
\be
p'(r)\equiv\frac{dp}{dr}=-\frac{M(r) \rho(r)}{8\pi r^2}\,, \label{hydrostatic}
\ee
with $p=\rho=0$ for $r \ge r_*$ and 
\be
M(r)=4 \pi \int_0^{r} \rho({\tilde r}) {\tilde r}^2 d {\tilde r}\,.
\ee

Taking into account that
\begin{equation}
\langle p \rangle = \frac{\int_0^{r_*} p(r) r^2 dr}{\int_0^{r_*} r^2 dr}=-\frac{\int_0^{r_*} p'(r) r^3 dr}{3\int_0^{r_*} r^2 dr}\,,
\end{equation}
and assuming, for the sake of definiteness, that $\rho(r)={\langle \rho \rangle}=3M_*/(4\pi r_*^3)$, the average pressure $\langle p \rangle$ inside a compact object may be estimated as 
\begin{equation}
\langle p \rangle  \sim \frac{3}{160 \pi^2} \frac{M^2_*}{r_*^4}\,. \label{averagep1}
\end{equation}

Equation~(\ref{averagep1}) may also be written as 
\begin{equation}
\frac{\langle p \rangle}{\langle \rho \rangle} \sim  \frac{M_*}{40 \pi r_*}=\frac{|\phi_*|}{5}\,,
\end{equation}
where 
\be
\phi_*=-\frac{M_*}{8 \pi r_*}\label{phistar}
\ee
is the Newtonian gravitational potential at the surface of the object. $|\phi_*|$ can vary by many orders of magnitude when one considers objects with various masses and sizes. For example,
\begin{equation}
|\phi_{\odot}| \sim 2 \times 10^{-6}\,,   \quad |\phi_{pr}| \sim 10^{-39}\,,
\end{equation}
where $\odot$, and $pr$ represent the sun and the proton, respectively. Here, we have taken the following values, in units of the International System, for the masses and radius of these objects: $m_\odot=2.0 \times 10^{30} \, \rm kg$, $r_\odot=7.0 \times 10^{8} \, \rm m$, $m_{pr}=1.7 \times 10^{-27} \, \rm kg$, and $r_{pr} \sim 10^{-15} \rm m$. 

In general relativity the condition $\langle p \rangle/\langle \rho \rangle = 0$ is approximately verified, both for the sun  and the proton, but specially in the later case. This is to be expected since in general relativity the gravitational field inside the proton is extremely weak so that the spacetime is essentially Minkowskian.  Nevertheless, as it will be shown in Sec. \ref{sec:eibi}, there are viable modified theories of gravity in which the intensity of the gravitational field can be enhanced by many orders of magnitude on microscopic scales with respect to general relativity.  These theories may therefore be probed more effectively through experiments which are sensitive to interactions on such small scales.

\section{Effective gravitational pressure in EiBI gravity  \label{sec:eibi}}

In this section we shall briefly describe the main features of EiBI gravity, giving particular relevance to the concept of effective gravitational pressure. 

The EiBI theory of gravity is described by the action
\be
S=\frac{2}{\kappa}\int d^{4}x\left[\sqrt{\left|{\rm det}(g_{\mu\nu}+\kappa R_{\mu\nu})\right|}-\lambda\sqrt{|g|}\right]+S_M\,,\label{eq:EddingtonBornInfeld Action}
\ee
and it is based on the Palatini formulation which treats the metric and the connection as independent fields.
Here, $g_{\mu\nu}$ are the components of the physical metric, $g \equiv {\rm det}(g_{\mu\nu})$ is the determinant of $g_{\mu\nu}$, $R_{\mu\nu}$ is the symmetric Ricci tensor build from the connection  $\Gamma$, $S_M$ is the standard action associated with the matter fields, and $\kappa$ is the only additional parameter of the theory with respect to general relativity. Without loss of generality we set $\lambda=1$ (the changes associated with a different value of $\lambda$ can be incorporated into the energy-momentum tensor).

The equations of motion,
\bq
q_{\mu\nu}&=&g_{\mu\nu}+\kappa R_{\mu\nu}\,,\label{eq:ConnectionEquationOfMotion}\\
\sqrt{|q|}q^{\mu\nu}&=&\sqrt{|g|}g^{\mu\nu}-\kappa\sqrt{|g|}T^{\mu\nu}\,,\label{eq:MetricEquationOfMotion}
\eq
may be derived by varying the action with respect to the connection and the physical metric. Here, $T^{\mu \nu}$ are the components of the energy-momentum tensor, $q_{\mu\nu}$ is an auxiliary (apparent) metric related to the original connection by 
\be
\Gamma^{\gamma}_{\mu\nu} = {1 \over 2} q^{\gamma\zeta}(q_{\zeta\mu,\nu} + q_{\zeta\nu,\mu}- q_{\mu\nu,\zeta})\,, \label{connection}
\ee
$q^{\mu\nu}$ is the inverse of $q_{\mu\nu}$, $q={\rm det}(q_{\mu\nu})$, and a comma represents a partial derivative. 

Combining Eqs.~(\ref{eq:ConnectionEquationOfMotion}) and (\ref{eq:MetricEquationOfMotion}) one obtains the second-order field equations
\be
{{\mathcal G}^\mu}_\nu \equiv {{{\mathcal R}}^\mu}_\nu -\frac12 {\mathcal R} {\delta^\mu}_\nu  ={{\mathcal T}^{\mu}}_{\nu}\,,\label{eq:EquationOfMotionComb1}
\ee
with 
\bq
{{{\mathcal R}}^\mu}_\nu &\equiv& q^{\mu \zeta} R_{\zeta \nu} ={\Theta^{\mu}}_{\nu}\,,\label{eq:EquationOfMotionComb}\\
{{\mathcal T}^{\mu}}_{\nu} &\equiv& {\Theta^{\mu}}_{\nu}-\frac12\Theta {\delta^\mu}_\nu\,,\label{eq:aThetamunu}\\
{\Theta^{\mu}}_{\nu}&\equiv&\frac{1}{{\kappa}}\left(1-\tau\right){\delta^\mu}_\nu+\tau{T^{\mu}}_{\nu} \,,\label{eq:Thetamunu}\\
\Theta &\equiv&  {\Theta^{\mu}}_{\mu}\,\label{eq:Theta}\\
\tau&\equiv&\sqrt{{\frac{g}{q}}}= \left[ \det( {\delta^\mu}_\nu -  \kappa {T^\mu}_\nu ) \right]^{-\frac{1}{2}} \label{eq:tau}\,.
\eq
Here, ${{\mathcal G}^\mu}_\nu$ are the components of the apparent Einstein tensor (analogous to the physical Einstein tensor of general relativity, with the physical metric replaced by the apparent metric), ${{\mathcal T}^{\mu}}_{\nu}$ are the components of the apparent energy-momentum tensor, and ${\delta^\mu}_\nu$ is the kronecker delta.

The components of the energy-momentum tensor of a perfect fluid are given by
\begin{equation}
{T^{\mu}}_{\nu}=(\rho+p)u^\mu u_\nu + p {\delta^{\mu}}_{\nu}\,,
 \end{equation}
where $\rho$ and $p$ are the proper physical energy density and pressure, respectively, and $u^\mu$ are the components of the 4-velocity of the fluid ($u^\mu u^\nu g_{\mu \nu} =-1$). If the source of the gravitational field is a perfect fluid then $\tau$ is given by
\be
\tau=\left[ (1+\kappa \rho)(1-\kappa p)^3 \right]^{-\frac{1}{2}}\, \label{eq:tau1}
\ee
In \cite{Delsate:2012ky} it has been shown that it is possible to write the apparent energy-momentum tensor in a perfect fluid form
\begin{equation}
{{\mathcal T}^{\mu}}_{\nu}=(\rho_q+p_q)v^\mu v_\nu + p_q {\delta^{\mu}}_{\nu}\,,
 \end{equation}
where $v^\mu$ are the components of the apparent 4-velocity of the fluid (satisfying $v^\mu v^\nu q_{\mu \nu}=-1$), and the proper apparent  pressure and energy density are given, respectively,  by
\bq
p_q &=& \tau p + \mathcal{P}\,, \label{eq:pqrq}\\
\rho_q &=&\tau \rho - \mathcal{P}\,, 
\eq
with
\be
\mathcal{P} = \frac{1}{\kappa}\left(\tau - 1\right) - \frac{1}{2} \tau \left(3p-\rho\right) \label{eq:P}\,. 
\ee
Hence, the apparent pressure is just the sum of the physical pressure with an additional effective gravitational pressure contribution $p_G$ defined by
\be
p_G=p_q-p\,.
\ee
Expanding \eqref{eq:pqrq} up to first order in $\kappa \rho$ and $\kappa p$, using also Eqs. \eqref{eq:tau1} and Eq. \eqref{eq:P}, one obtains that
\be
p_G=p_q-p= \kappa p^2+\frac{\kappa}{8} \left(\rho+ p\right)^2\,, \label{eq:pgp1}
\ee
where $p_G$ is always positive for $\kappa >  0$ and always negative for $\kappa < 0$.

\subsection{Newtonian limit of EiBI gravity}

Consider a time independent metric in the Newtonian gauge described by the line element
\be
ds^2=-(1+2\Phi)dt^2 + (1-2\Psi) \delta_{\mu \nu} dx^\mu dx^\nu\,. \label{newtonian}
\ee
In the Newtonian limit of EiBI gravity the equation of motion for the gravitational field is the usual Poisson equation with an extra source term \cite{Banados:2010ix}:
\be
\nabla^2 \Phi = \frac{\rho}{2} + \frac{\kappa}{4} \nabla^2 \rho\,, \label{Poisson}
\ee
Note that $\Phi=\Psi$ and $p \ll \rho$ in the Newtonian limit. Equation~\eqref{Poisson} may be cast in the usual form, 
\be
\nabla^2\phi=\frac{\rho}{2}\,,\label{Poisson_q}
\ee
by defining the apparent gravitational potential as
\be
\phi \equiv \Phi-\kappa \frac{\rho}{4}\,.
\ee
The gravitational acceleration is given by
\be
{\vec a}=-\nabla\Phi=-\nabla\phi-\frac{\kappa}{4}\nabla\rho =-\nabla\phi -\frac{1}{\rho}\nabla p_G \,. \label{acceleration}
\ee
The second term on the right hand side of Eq.~\eqref{Poisson} may now be interpreted as the source of an acceleration associated to the effective gravitational pressure contribution $p_G$ which, in the Newtonian limit, is just given by $p_G=\kappa \rho^2/8$. This term may be responsible, if $\kappa \neq 0$, for a strong gravitational acceleration field associated to small scale variations of the density field (see \cite{Avelino:2012ge} for a complementary description in terms of a constant effective Jeans length).

 \section{Sub-femtometer constraints on gravity}
 \label{sec:proton}

In this section we shall investigate the constraints on modified theories of gravity which can be derived from measurements of the proton interior pressure profile, paying special attention to the case of EiBI gravity. We shall consider, in particular, the quark pressure measurement reported in \cite{Burkert:2018bqq}, and discuss the improvements which will result from more precise measurements of the pressure distribution inside the proton. The discussion in Sec. \ref{sec:gr} implies that, in general relativity, the gravitational potential inside the proton is tiny, and may, therefore, be neglected. On the other hand, in Sec. \ref{sec:eibi} it was shown that EiBI gravity  essentially reduces to general relativity with an apparent metric field sourced by an apparent energy-momentum tensor --- the  proper apparent energy density and pressure being of the same order or smaller than the proper physical energy density, except if $\tau$ deviates significantly from unity. Hence, the apparent metric inside the proton in EiBI gravity is essentially the Minkowski metric and the von-Laue condition given in Eq. (\ref{averagep}) holds if the physical pressure $p$ is replaced by the total pressure
\be
p_T=p+p_G\,. \label{eq:p1}
\ee
Therefore,
\be
\langle p_T \rangle = 0\,, \label{eq:p2}
\ee 
and the condition
\be
|\langle p \rangle| =  |\langle p_G \rangle|\,, \label{eq:p3}
\ee
is always an excellent approximation. Hence, a constraint on the magnitude of the average pressure $p$ inside the proton also represents a constrain on the magnitude of the average effective gravitational pressure $p_G$. 

Let us now consider the particular case of EiBI gravity. In combination with the dominant energy condition \cite{1973lsss.book.....H} (which implies that $\rho \ge |p|$), Eq. \eqref{eq:tau} shows that the constraint $|\kappa|<|p|^{-1}$ must be always satisfied in order that $\tau$ remains finite. Hence, the measurement of an average peak pressure around $10^{35} \, \rm Pa$, near the center of the proton, may be translated into the following upper limit on $|\kappa|$
\be
|\kappa| \lsim c^4 |p_{\rm peak}|^{-1} \sim 10^{-1} \, {\rm m^5 \, kg^{-1} \, s^{-2}}\,. \label{eq:kl1}
\ee
This limit is roughly one order of magnitude stronger than that obtained from the predicted core density of neutron stars \cite{Pani:2011mg} (note that we explicitly included the factor of $c^4$ in Eq.  \eqref{eq:kl1} before evaluating the upper bound on $|\kappa|$ in units of the International System).

We will now show that the limit given in Eq. \eqref{eq:kl1} may be significantly improved with precise measurements of the first and second moments of the pressure distribution inside the proton ($\langle p \rangle$ and $\langle p^2 \rangle$, respectively). Using Eqs. \eqref{eq:aThetamunu} to \eqref{eq:tau}, it is possible to show that the apparent pressure is equal to
\bq
p_q &=& \frac{{{\mathcal T}^{i}}_{i}}{3}=\frac{1}{\kappa}\left(\tau - 1\right) + \tau \left( \frac{{{T}^{i}}_{i}}{3} - \frac{{T^{\mu}}_{\mu}}{2}   \right) \nonumber\\
&=&\frac{1}{\kappa}\left(\tau - 1\right) + \tau \frac{\rho-p}{2}\,, \label{eq:pqnew}
\eq
where $p \equiv {{{T}^{i}}_{i}}/{3}$ and ${T^{\mu}}_{\mu}=-\rho+3p$, for any fluid at rest (not necessarily perfect).

The energy-momentum tensor of the proton is not that of a perfect fluid. If the proton were a spherically symmetric spin-0 particle, the non-zero components of the energy-momentum tensor would be 
\be
{T^0}_0=-\rho\,, \quad {T^r}_r=p_r(r)\,, \quad {T^\theta}_\theta={T^\phi}_\phi=p_\perp(r)\,, \label{eq:pem1}
\ee
in a static frame where the proton is at rest (here, $[r,\theta,\phi]$ are spherical coordinates). In this case, Eqs. \eqref{eq:pqrq} and \eqref{eq:P} would remain valid, but $\tau$ would now given by
\be
\tau=\left[ (1+\kappa \rho)(1-\kappa p_r) (1-\kappa p_\perp)^2\right]^{-\frac{1}{2}}\,. \label{eq:tau2}
\ee
Expanding Eq. \eqref{eq:pqnew} up to first order in $\kappa \rho$, $\kappa p_r$ and $\kappa p_\perp$, using also Eqs. \eqref{eq:pem1} and \eqref{eq:tau2}, one obtains that
\be
p_G=p_q-p= \kappa p^2+\frac{\kappa}{8} \left(\rho+ p\right)^2 +\frac{\kappa}{6}\left(p_r-p_\perp\right)^2\,, \label{eq:pgp1}
\ee
where $p=(p_r+2 \, p_\perp)/3$, and $p_G$ is always positive for $\kappa >  0$ and always negative for $\kappa < 0$. Therefore, 
\be
|p_G| \ge |\kappa| p^2 \,. \label{eq:pgp2}
\ee
Using Eqs. \eqref{eq:tau} and \eqref{eq:pqnew} it is possible to show that Eq. \eqref{eq:pgp2} holds (up to first order in $\kappa {T^\mu}_\nu$) even in the presence of off-diagonal contributions to the energy-momentum tensor, such as those associated to the spin of the proton --- these off-diagonal contributions would only add new terms of the form $\kappa ({T^\mu}_\nu)^2/2$, with $\mu \neq \nu$,  to Eq. \eqref{eq:pgp1}.

Equations  \eqref{eq:pgp1} and  \eqref{eq:pgp2} also imply that
\be
| \langle p_G \rangle|  = | \langle p \rangle| \ge  |\kappa| \langle p^2 \rangle\,,
\ee
or, equivalently, that
\be
|\kappa|  \le   \frac{ | \langle p \rangle|}{ \langle p^2 \rangle} = \frac{\xi}{\sqrt{\langle p^2 \rangle}}\,, \label{eq:kconst}
\ee
where the dimensionless parameter 
\be
\xi\equiv\frac{\langle p \rangle}{\sqrt{\langle p^2 \rangle}}\,,
\ee 
represents a quantitative measure of the impact of gravity on the interior structure of the proton.

Taking into account that $\langle(p-\langle p \rangle)^2 \rangle = \langle p^2 \rangle - \langle p \rangle^2 \ge 0$, the condition
\be 
\xi\equiv {\langle p \rangle}/{\sqrt{\langle p^2 \rangle}} \le 1
\ee
is always verified. Hence, Eq. \eqref{eq:kconst} implies that $|\kappa|  \le  1/\sqrt {{\langle p^2 \rangle}}$. However, this is always a weaker constraint than $|\kappa|  \le   |p_{\rm peak}|^{-1}$ (given in Eq. \eqref{eq:kl1}). Therefore, in order to be able to take full advantage of Eq. \eqref{eq:kconst}, both the first and the second moments of the pressure distribution inside the proton need to be measured with sufficient precision.

Here we have followed Ref. \cite{Burkert:2018bqq} and implicitly assumed that the radial profile of the gluon pressure --- the other major contribution to the pressure inside the proton  --- is similar to that of the protons. Future experiments, such as the Electron Ion Collider \cite{Accardi:2012qut} should be able to test this assumption, making it possible to obtain more direct and precise estimates of the total pressure distribution inside the proton. In \cite{Burkert:2018bqq} the authors report that Eq. \eqref{averagep} is satisfied within the uncertainties of the measurement of the quark  pressure distribution inside the proton. However, present uncertainties on the radial quark pressure profile are still large \cite{Burkert:2018bqq}, not allowing for the use of the relation provided in Eq. \eqref{eq:kconst} to significantly improve upon the constraint given in Eq. \eqref{eq:kl1} (even assuming similar contributions from quarks and gluons to the pressure inside the proton). Nevertheless, more precise reconstructions of the quark and gluon pressure distribution inside the proton, which will be made possible in the future  \cite{Burkert:2018nvj,Accardi:2012qut}, are expected to yield useful constraints on $\xi$ and, therefore, to lead to stronger limits on modified gravity theories (including tighter constraints on the EiBI parameter $\kappa$).

In \cite{Latorre:2017uve} an extremely stringent constraint on metric affine gravity (including EiBI gravity as a particular model) was found by requiring consistency between Bhabha scattering experimental results and the corresponding classical scattering cross-sections computed at tree level in EiBI gravity. These results should be taken with some caution, since they do not account for loop corrections. Also, note that the quantization of the matter fields in the context of EiBI gravity is subtle (in particular, since the physical metric may be significantly distorted even if the apparent metric is essentially Minkowskian).

Here, we have conservatively assumed that EiBI gravity does not affect the determination of the physical pressure distribution inside the proton using virtual Compton scattering. Therefore, the limits on EiBI gravity discussed in the present paper should be regarded as conservative bounds.

\section{Conclusions}\label{sec:conc}

In this paper we have shown that the recent measurement of the pressure distribution experienced by the quarks inside the proton may be used to constrain modified theories of gravity in which the strength of the gravitational field on microscopic scales is enhanced with respect to general relativity. We applied these results to the particular case of EiBI gravity, showing that strong limits on the single additional parameter of theory with respect to general relativity $\kappa$ may be derived from the measurement of the interior pressure profile of the proton. We have also shown how these limits may be significantly improved  with precise measurements of the first and second moments of the pressure distribution inside the proton. 
	
\begin{acknowledgments}

We thank Adrià Delhom-Latorre for enlightening discussions. This work was supported by FCT - Fundação para a Ciência e a Tecnologia through national funds (PTDC/FIS-PAR/31938/2017) and by FEDER - Fundo Europeu de Desenvolvimento Regional through COMPETE2020 - Programa Operacional Competitividade e Internacionalização (POCI-01-0145-FEDER-031938). Funding of this work has also been provided by FCT/MCTES through national funds (PIDDAC) by the grant UID/FIS/04434/2019.

\end{acknowledgments}
 
\bibliography{proton}
 	
 \end{document}